\documentclass{edp-conf}
\usepackage{graphicx}
\begin{document}

\TitreGlobal{SF2A 2005}

\title{The Galaxy in the Cosmological Context}
\author{Combes, F.}\address{Observatoire de Paris, LERMA, 61 Av. de l'Observatoire, F-75014, Paris}
\runningtitle{The Galaxy in the Cosmological Context}
\setcounter{page}{237}
\index{Combes, F.}
%
\maketitle
\begin{abstract}
  A new view on our Galaxy has recently emerged, with large consequences on its formation scenarios. Not only new dwarf satellites 
have been detected, still orbiting and tidally disrupting, but also a multitude of stellar streams or tidal debris have been observed 
suggesting the formation of the halo through successive accretions. The large scatter in the age-metallicity relation in the solar
 neighborhood points towards several accretion episodes, while the chemical evolution of the disk requires a more or less continuous 
gas infall. The global kinematics and morphology refined by large surveys such as 2MASS suggest the existence of two embedded 
bars, as is frequently observed in external galaxies. The mass of the central black hole has been refined through stellar proper motions, 
and is compatible with the M$_{bh}$-$\sigma$ relation valid for all bulges. The baryonic dark matter is no longer thought to lie in 
compact objects, and on the contrary, more dark cold gas is revealed by gamma-ray observations. The star forming history can be built, 
and confronted to numerical models of galaxy evolution both through hierarchical and secular scenarios. Our Galaxy plays thus the 
role of a prototype to probe galaxy formation theories, and in particular thin and thick disk formation.
\end{abstract}
%
\section{Chemical evolution}
  Evolution can be probed in our Galaxy through a detailed 
study of stellar populations, and in particular the large program
on VLT-UVES (e.g. Cayrel et al 2004) has unveiled suprising 
characteristics of the most metal-poor stars, which are certainly
the oldest in the Galaxy.  There is very little variations
of relative abundances [X/Fe], implying a good mixing of old stars
(Fran\c{c}ois et al 2004).  The N and N/O ratios, compatible
with those found in DLA systems, could come either
from the first SNe II or from massive AGB stars (and well before
any enrichment through SNe Ia, Spite et al 2005).
 The constraints given by the abundances, and the ionization epoch
tend to imply that the first IMF was top-heavy, and there were no small-mass
Pop III stars. Discrepant results on Cu, Co and Zn abundances put into
questions the yields (implying for instance asymmetric supernovae explosions,
e.g. Prantzos 2005).
 
A new unbiased survey of 14 000 F \& G dwarfs in the solar neighborhood,
with age, metallicity and kinemactical parameters, has shown that there is
no age-metallicity relation, with a large Z-scatter, for all ages except the
very young (Nordstr\"om et al 2004). The survey confirms the G-dwarf problem, 
and the radial gradient of metallicity, for young stars ($<$ 10 Gyr). The 
confirmed age-velocity dispersion relation is explained by the scattering
from GMC and spiral arms.  From this survey, Helmi et al (2005) 
have identified a lot of structures in the solar neighborhood,
most of them coming from spiral arms, bar, etc.. and some
coming from tidal debris.

\section{Halo streams}

Many tidal debris and streams have been recently discovered in 
the stellar halo, suggesting that it could have been entirely formed
from accreted satellites (e.g. Ibata et al  2003). The same situation 
has been found around M31 (Ibata et al 2004). The streams from
SgrA dw could serve to determine the flattening of the dark
matter halo, at least those which are old enough (Helmi 2004).
Willman et al (2005) have found recently another very weak possible
satellite in Ursa Major, much weaker than Sextans or any other dwarfs.

\section{Bars, spirals and central black hole}

After the successful model of 4-arm spiral pattern (Georgelin \& Georgelin 1976)
some more precision about spiral arms is found by Russeil et al (2005).
  The central bar-bulge (and may be secondary bar? Alard 2001) has been explored
in details with the 2MASS survey (cf Figure 1, and Lopez-Corredoira
et al 2005).

\begin{figure}[h]
   \centering
   \includegraphics[width=9cm]{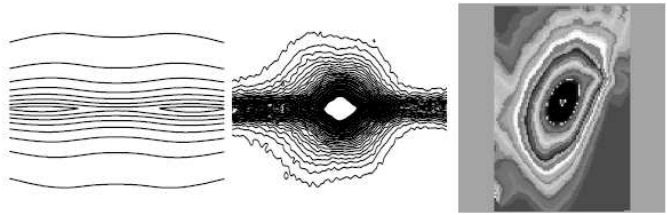}
      \caption{2MASS star counts of the central Milky Way.
Left: Star counts of the disc only contribution;
Middle:  Total star counts when the disc contribution has been
subtracted;  Right: Cut at $z=-440$ pc.
From Lopez-Corredoira et al (2005).
}
       \label{fig1}
   \end{figure}

Our Galaxy is a privileged example to study 
central black holes in galaxies, through proper motions
(Sch\"odel et al 2003, Ghez et al 2005).  Flares have been observed in 
near-infrared, that could be emitted by matter
on the last stable orbit around the black hole, and could reveal
its rotation (Genzel et al 2003).

\section{Dark matter}

More stellar tracers of the kinematics in the outer halo of the Milky Way have
recently been obtained by Battaglia et al (2005).  
The radial velocity dispersion declines from 120 km/s to 50 km/s,  from 30 kpc to about 120 kpc,
which puts constraints on the density profile and total mass of the dark matter. 
 These depend  on the velocity anisotropy assumed, but an isothermal profile
is now excluded, and the total mass  somewhat reduced (cf Figure 2, Battaglia et al 2005).

\begin{figure}[h]
   \centering
   \includegraphics[width=8.5cm]{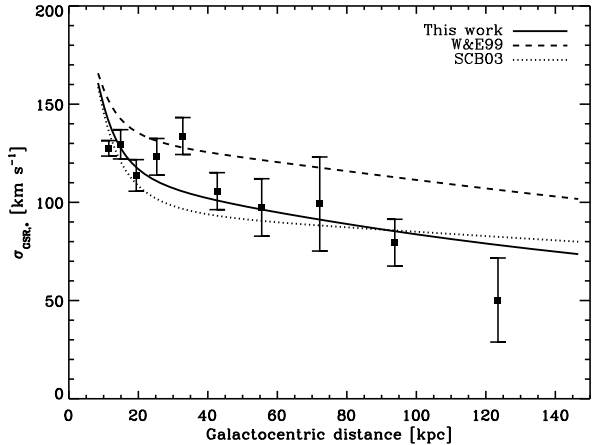}
      \caption{Observed radial velocity dispersion (squares with error-bars)
overlaid on the best fit model of TF (Truncated Flat) mass distribution (solid
line). The dashed and dotted lines correspond to
the best-fitting parameters from 
Wilkinson \& Evans, 1999 (dashed) and Sakamoto et al, 2003 (dotted).  
From Battaglia et al (2005).
}
       \label{fig2}
   \end{figure}

As for the local dark matter, Grenier et al (2005) have recently found
evidence for a large amount of dark cold gas, which could be in the form
of molecular hydrogen. This result is based on the observation of energetic gamma-rays
diffuse emission, resulting from the interaction between nucleons and cosmic rays.
The emission is associated to excess extinction, around known interstellar
clouds. Most of it was previously attributed to point sources (pulsars),
associated to local clouds, which have never been confirmed.

The Milky Way is not the ideal galaxy to constrain the amount and radial 
distribution of the dark matter, since the inner regions are completely
dominated by visible matter. Models by Klypin et al (2002) already 
succeeded to find reasonable fits, with more or less baryons or dark matter
in the center, according to the amount of angular momentum lost
by baryons against the CDM halo. But the dark matter never dominates in 
the central parts, and there must exist dark baryons within the virial
radius of the halo. Recently Cardone \& Sereno (2005) built a
larger range of models, with the hypothesis of adiabatic contraction,
varying the parameters of the DM distribution. Their best fit
corresponds to much lower concentration of dark matter profiles.

\section{Star formation history}

The chemical evolution observed in the stellar populations, together
with other constraints, have allowed to reconstruct the star formation
history of the Milky Way. For instance, in the disk, the star formation rate (SFR)
has remained relatively constant, or is even increasing (Haywood et al 1997).
 Abundance problems (such as the G-dwarf problem) require large
infall of external gas to dilute enrichment (Casuso \& Beckman 2004).
Naab \& Ostriker (2005) have tried to reproduce the star formation history
all across the Hubble time, assuming that the halo assembles through 
hierarchical merging of subhalos until 2.5 Gyr, and then the disk is
forming inside out, by only quiescent accretion of gas. 
Adopting a simple Schmidt law with threshold for the SFR,
the resulting model fits the metallicity and mean age distribution
of the Galaxy, with an SFR peaking about 4 Gyr ago. 


\end{document}